%Itzhak Bars, Strings and matrix models on genus g Riemann Surfaces
%hep-th/9706177, USC-91/HEP-B1.
\overfullrule=0pt
  
%%%%%%%%%%%%%%%%%%%%%%%%%%%%%%%%%%%%%%%%%%%%%%%%%%%%%%%%%%%%%%%%%

% ONLY PLAIN TEX IS NEEDED TO TEX THIS FILE. THERE ARE NO MACROS. 

%%%%%%%%%%%%%%%%%%%%%%%%%%%%%%%%%%%%%%%%%%%%%%%%%%%%%%%%%%%%%%%%%
\def\no{\noindent}
\def\vn{\vec n}
\def\vm{\vec m}

%%%%%%%%%%%%%%%%%%%%%%%%%%%%%%%%%%%%%%%%%%%%%%%%%%%%%%%%%%%%%%%%%%%%

%\font\fone=amr10 scaled\magstep1
%\font\ftwo=amr7 scaled\magstep0
%\font\fthree=amr10 scaled\magstep0

%%%%%%%%%%%%%%%%%%%%%%%%%%%%%%%%%%%%%%%%%%%%%%%%%%%%%%%%%%%%%%%%%
\nopagenumbers 
\hsize=37pc \vsize=50pc \parskip=1pc \parindent=25pt \baselineskip=12pt 
\magnification=1200

%\input {\tex\harvmac}

%%%%%%%%%%%%%%%%%%%%%%%%%%%%%%%%%%%%%%%%%%%%%%%%%%%%%%%%%%%%%%%%%%%

\noindent{USC-91/HEP-B1} \hfill {hep-th/9706177}
\break\noindent{January 1991}
\bigskip\bigskip

\centerline {\bf Strings and matrix models on genus g Riemann Surfaces}
\medskip
\centerline {  {old title:\bf STRINGS AND LARGE N QCD}
            {  \footnote {$^*$}  
            {   Research supported in part by the 
              U.S. Department  of Energy, under Grant No. 
              DE-FG03-84ER-40168    }   }  
             }

\bigskip
\centerline {Published in the proceedings of ``Beyond the Standard Model", Ed. K. Milton, 1991}

\vskip 1.00 true cm

\centerline { ITZHAK BARS }

\bigskip

\centerline {Physics Department, University of Southern California}
\centerline {Los Angeles, CA 90089-0484, USA}

%%%%%%%%%%%%%%%%%%%%%%%%%%%%%%%%%%%%%%%%%%%%%%%%%%%%%%%%%%%%%%%%%%%%
\vskip 1.20 true cm
\centerline{ABSTRACT} \bigskip
 
  %   \hsize 39pc    \baselineskip=12pt  \parindent= 0.00 true cm
  %   \hangindent=6pc           %hanging indentation for entire par
  %   \hangafter=0                       %Type abstract on very next line

     \hsize 14.00 true cm
     \parindent= 0.00 true cm
     \hangindent=1.15 true cm           %hanging indentation for entire par
     \hangafter=0                       %Type abstract on very next line
\bigskip
Comment: This is a summary of old work on connections between discrete area preserving diffeomorphisms, reduced SU(N) Yang-Mills, strings, and the quantum Hall effect on a Riemann surface of genus g. It is submitted to the archives due to the interest expressed by colleagues who are currently working on matrix models, and who could not have access to the proceedings in which the article was published. The text that follows is the version published in 1991.
\hfill\break 
.\hfill\break 
In this talk I will describe an attempt to bridge string theory and large N 
QCD, as obtained in  recent papers. This is based on a relation between 
area preserving diffeomorphisms and $SU(\infty)$. The reduced model of QCD 
takes the form of a version of string theory that is related to ordinary 
string theory in the gauge $det (\gamma)=-1$, where $\gamma_{ij}$ is the world 
sheet metric.

     \parskip=0.25 true cm       %adjust to taste
     \hsize=32pc  \baselineskip=12pt  \parindent=20pt  %repair the damage

%%%%%%%%%%%%%%%%%%%%%%%%%%%%%%%%%%%%%%%%%%%%%%%%%%%%%%%%%%%%%%%%%%%%
\bigskip
\no
{\bf 1. Area Preserving Diffeomorphisms and $SU(\infty)$}

Consider a compact Riemann surface parametrized by the Euclidean parameters 
$\sigma^i=(\tau,\sigma)$. The infinitesimal transformations 
$\sigma^i\rightarrow \sigma^i+\xi^i$ that leave the area element 
$d^2\sigma=d\tau d\sigma$ invariant are the area preserving diffeomorphisms. 
The parameters for such transformations satisfy $\partial_i\xi^i=0$ and can be 
expressed in the form  $\xi^i=\epsilon^{ij}\partial_j\Omega + \sum_r
\lambda_rv_r^i$, where $\Omega(\tau,\sigma)$ is an arbitrary function (except 
for periodicity properties on the surface), $\lambda_r$ are constants and 
$v_r^i(\tau,\sigma)$ are the harmonic 1-forms, of which there are as many as 
twice the genus of the surface, $r=1,2,\cdots 2g$. Corresponding to $\Omega$ 
we have the local subalgebra whose generator is labelled as $L(\tau,\sigma)$ 
and corresponding to $\lambda_r$ we have the global translation operators
$K_r$ which generate translations along the $2g$ cycles on the surface. The 
algebra of these generators was given for the sphere [1] and torus [2]. The 
general commutation rules for any Riemann surface of genus $g$ including the 
general form of a potential anomaly, was obtained in [3], where 
generalizations to supersymmetry and higher dimensions was also given. Here we 
will  mainly concentrate on the local subalgebra of $L(\tau,\sigma)$. 

The relation to $SU(\infty)$ was explicitly given on the sphere [1] and torus 
[2] and it has only recently been obtained for surfaces of any genus 
[4] as will be described here. First consider the torus. We expand in terms 
of a complete set of periodic functions on the torus, one writes 
$L(\vec\sigma)= \sum_{\vec n} L_{\vec n}\ exp(i\vec n\cdot \vec\sigma)$, with 
$\vn=(n,n')$, and obtain the Lie algebra of area preserving diffeomorphisms in 
Fourrier space as 

$$ [ L_{\vec n}, L_{\vec m}] = i (\vec n \times \vec m) L_{\vec n+\vec m} 
\eqno(1) $$

\no
where $(\vec n\times \vec m)=nm'-mn'$. This algebra is related to 
SU(N) as $N\rightarrow \infty$ as follows.
Consider the $N\times N$ (N=odd) Weyl matrices $h$ and $g$ that satisfy 
$h^N=1=g^N$ and $gh=hg\omega$, where $\omega=exp(i4\pi/N)$. These matrices 
are explicitly given as $h=diag (1,\omega,\omega^2,\cdots,\omega^{N-1})$ and 
$g_i^j=\delta_{i+1}^j$ defined by identifying the 
indices $i\ or j=N+1\rightarrow 1$, so that it has non-zero entries only above 
the diagonal and at the $(i,j)=(N,1)$ location. There are $N^2$ linearly 
independent powers of these matrices, $h^{n}g^{n'},\ \ 
n,n'=0,1,\cdots,(N-1)$, that are unitary and close under multiplication. 
Excluding the $n=0=n'$ identity matrix, the remaining ones are traceless 
and close under commutation. Thus, we construct the SU(N) generators in this 
basis $ l_{\vec n}= {N\over 4\pi} h^{n}g^{n'} \omega^{nn'/2}$. They can 
be shown to satisfy the commutation rules

$$ [l_{\vec n}, l_{\vec m}] =i {N\over 2\pi} sin ({2\pi\over N}\vec n\times 
\vec m) \ l_{\vec n + \vec m} ,  \eqno(2) $$

\no
that parallels (1). In this form, taking the $N\rightarrow \infty$ limit 
(1) and (2) become identical, thus displaying the relation between area 
preserving diffeomorphisms and $SU(\infty)$ on the torus. Note, however, that 
this is true only if $(n\times m)/N$ is small, which means the relationship 
between $SU(\infty)$ and area preserving diffeomorphisms can be valid only 
when integrated with an appropriate set of functions. This is analogous to the 
equivalence between classical mechanics and quantum mechanics in the limit of 
$\hbar\rightarrow 0\ \sim ({1\over N} \rightarrow 0)$, provided we use an 
appropriate set of wavefunctions. A similar construction on the sphere uses 
the spherical harmonic basis $Y_{jm}$ [1]. 

For a genus $g$ surface, following ref.[4] we consider $SU(N)$ with 
$N=N_1\times N_2\times\cdots N_g$. We label the $N$-dimensional fundamental 
representation by a composite index $\psi_{i_1i_2\cdots i_g}$ where 
$i_1=1,2,\cdots N_1;\ i_2=1,2,\cdots N_2;\ i_g=1,2,\cdots N_g$. We see that we 
can construct the subgroups $SU(N_1),\ SU(N_2), \cdots , SU(N_g)$ in a direct 
product basis 

$$ l_{\vn_1}^{(N_1)}\times 1_{N_2}\times \cdots \times 1_{N_g} \qquad
 1_{N_1}\times l_{\vn_2}^{(N_2)} \cdots \times 1_{N_g} \qquad
 1_{N_1}\times 1_{N_2}\times \cdots \times l_{\vn_g}^{(N_g)}   \eqno(3) $$

\no
where $l_{\vn_k}^{(N_k)} $ is a $N_k\times N_k$ matrix constructed from $g,h$ 
matrices of rank $N_k$ and satisfies (2). We can then construct the full 
$SU(N)$ by taking all possible $N\times N$ matrices of the direct product form 
($C_N$ is a constant, see below)

$$ l_{\vn_1\vn_2\cdots\vn_g}=C_N\ \left (h^{n_1}g^{n'_1}\times h^{n_2}g^{n'_2}\times 
\cdots \times h^{n_g}g^{n'_g} \right )\ exp\big ( i2\pi\sum_i{n_in'_i\over N_i} \big) \eqno(4) $$

\no
We can show that these $N\times N$ matrices satisfy the following matrix 
product rules 

$$ l_{\vn_1\vn_2\cdots \vn_g}\ l_{\vm_1\vm_2\cdots \vm_g}= C_N \ exp \big ( 
i2\pi\sum_{i=1}^g {\vn_i\times \vm_i\over N_i}\big ) \ 
l_{\vn_1+\vm_1,\vn_2+\vm_2,\cdots,\vn_g+\vm_g} \eqno(5) $$
   
\no                                                 
and the commutation rules 

$$ [l_{\vn_1\vn_2\cdots \vn_g}, l_{\vm_1\vm_2\cdots \vm_g}]= 2iC_N\ sin \big ( 
i{2\pi}\sum_{i=1}^g {\vn_i\times \vm_i\over N_i}\big ) \ 
l_{\vn_1+\vm_1,\vn_2+\vm_2,\cdots,\vn_g+\vm_g} \eqno(6) $$

\no
which generalize the $SU(N)$ commutation rules of (2) to arbitrary genus.

It can be shown [4] that the $SU(N)$ transformations generated by 
$l_{\vn_1\vn_2\cdots \vn_g}$ are closely related to discrete area preserving 
transformations of a lattice Riemann surface of genus $g$. This connection is 
obtained through generalized Jacobi theta functions defined on the genus $g$ 
surface. A particular set of these functions is identified with our labeling 
of the $N$-dimensional fundamental representation above

$$ \psi_{i_1i_2\cdots i_g}= Theta {{i_1\over N_1} {i_2\over 
N_2}\cdots {i_g\over N_g}\brack 0\ 0\ \cdots \ 0} (\vec Z(z);\Omega)\times 
F(z,\bar z)  \eqno(7) $$ 

\no
where $z=\tau + i \sigma$ is a point on the Riemann surface, $\Omega_{ij}$ is 
the $g\times g$ period matrix, $Z_i(z)=\int_{z_0}^z \omega_i(z') dz'$ is the 
Jacobi variety and $\omega_i(z)$ are the Abelian differentials.  As is well 
known when integrated around the standard $\alpha_i,\beta_i$ cycles one has 
$\int_{\alpha_j}\omega_i=\delta_{ij},\ \int_{\beta_j}\omega_i=\Omega_{ij}$. We 
now make a lattice by dividing the $\alpha_j,\beta_j$ cycles into $N_j$ 
(not necessarily equal) intervals $\Delta\alpha_j,\Delta\beta_j$ such that the 
integration for each interval gives $\int_{\Delta\alpha_j}\omega_i={1\over 
N_i}\delta_{ij},\ \int_{\Delta\beta_j}\omega_{i}={1\over N_j}\Omega_{ij}$. 
This implies that when $z$ is translated by $n_j$ intervals along $\alpha_j$ 
and $n'_j$ intervals along $\beta_j$ we get a transformation on the Riemann 
surface of the form 

$$ z'=z+\sum_jn_j\Delta\alpha_j+\sum_jn'_j\Delta\beta_j,\qquad 
   Z_i(z')=Z_i(z)+{n_i\over N_i}+\Omega_{ij}{n'_j\over N_j} \eqno(8) $$

\no
Using the properties of the theta function we can now show explicitly that our 
$N$-dimensional basis undergoes the transformation

$$ \psi_{i_1i_2\cdots i_g}(z')= {1\over C} 
\left (l_{\vn_1\vn_2\cdots\vn_g}\right )_{i_1i_2\cdots i_g}^{j_1j_2\cdots j_g}
   \psi_{j_1j_2\cdots j_g}(z) \eqno(9) $$

\no
The factor $F(z,\bar z)$ in (7) is inserted to cancel the well known extra 
phase that appears in the transformation of the theta function when (8) is 
applied. We have thus demonstrated that translations on the $N^2$ points of 
the lattice Riemann surface are expressed by our $SU(N)$ generators given in 
eqs.(4-6,9). Taking linear combinations of all these translations with arbitrary 
continuous coefficients gives the full $SU(N)$.     

It is worth mentioning that, up to a $z$-dependent factor, our wavefunctions 
$\psi_{i_1i_2\cdots i_g}$ form the basis of linearly independent solutions to 
the problem of a charged particle moving on a Riemann surface of genus $g$ in 
the presence of a magnetic field [5]. The $SU(N)$ symmetry is then identified 
with the magnetic translation group. This provides an approach for studying 
the generalization of the quantum Hall effect problem for arbitrary Riemann 
surfaces [5]. This problem also connects to the solutions of topological 
quantum field theories in 2+1 dimensions.

We may now ask how to relate our original area preserving transformations 
$L(\tau,\sigma)$ to the matrices $l_{\vn_1\vn_2\cdots \vn_g}$ as $N\rightarrow 
\infty$? We postulate position-momentum like structures $Q_i(\tau,\sigma), 
P_i(\tau,\sigma)$ constructed on the Riemann surface of genus $g$, with 
Poisson brackets $\{Q_i,P_j\}=\partial_\tau Q_i\partial_\sigma P_j-
\partial_\sigma Q_i\partial_\tau P_j=\delta_{ij} $. For example for the 
torus $Q=\tau, P=\sigma$. Next we construct the basis functions 
$ f_{\vn_1\vn_2\cdots \vn_g}(\tau,\sigma)=\ exp(i\sum_in_iQ_i+n'_iP_i)$ which 
satisfy the Poisson brackets 

$$ \{f_{\vn_1\vn_2\cdots \vn_g},f_{\vm_1\vm_2\cdots 
\vm_g}\}=\sum_{i=1}^g(\vn_i\times\vm_i) \ 
f_{\vn_1+\vm_1,\vn_2+\vm_2,\cdots,\vn_g\vm_g}. \eqno(10) $$

\no
Then we can write 
$L(\tau,\sigma)=\sum L_{\vn_1\vn_2\cdots \vn_g} f_{\vn_1\vn_2\cdots \vn_g}$ 
where the operators $L_{\vn_1\vn_2\cdots \vn_g}$ satisfy commutation rules 
with the same structure constants as eq.(10), which is the $N\rightarrow\infty$ 
limit of eq(6) provided we choose $N_i=N^{1/g}$ and $C={N^{1/g}\over 
4\pi}$. This yields the generalization of eqs.(1,2) and of the discrete 
Riemann surface analysis given previously for the torus [6]. 

\bigskip
{\bf 2. Strings from large N QCD}

Large N theories including QCD may be analyzed by using reduced  Eguchi-Kawai 
models. Reduced models are known to reproduce the planar graphs. We have 
suggested that in a double scaling limit reduced QCD could describe the sum 
over all genus [6]. In the reduced model the original NxN matrix gauge field 
$(A_\mu)_i^j(x^\mu)$ is replaced by the same field at the single space-time 
point $x^\mu=0$.  Derivatives are replaced by a 
commutator with a fixed matrix $(P_\mu)_i^j$ that plays the role of the 
translation generator. The covariant derivative becomes 
$a_\mu=P_\mu+A_\mu\sim iD_\mu$. This leads 
to the Yang-Mills field strength 

$$(F_{\mu\nu})=[iD_\mu,iD_\nu]\rightarrow [a_\mu,a_\nu]_i^j\equiv 
(f_{\mu\nu})_i^j , \eqno(11) $$ 

\no
and the reduced gauge theory action and path integral [8,9,10]

$$        S_{red}=-{1\over 4}\Big({2\pi\over\Lambda}\Big)^d 
  {N\over g_d^2(\Lambda)}  Tr(f_{\mu\nu}f^{\mu\nu})  \qquad
\int \prod_\mu Da_\mu f(a_\mu)\ exp (iS_{red} (a))        \eqno(12)      $$

\no
where $\Lambda$ is a cutoff and $f(a)=\int \prod_\mu DU_\mu \delta (a_\mu-
U_\mu P_\mu U_\mu^\dagger )$, where $U_\mu$ is a unitary matrix corresponding 
to a reduced Wilson line integral [10].
                        
We wish to rewrite the large-N reduced model as a theory defined on Riemann 
surfaces reminiscent of string theory [6,7]. We illustrate this for the 
torus, however using the formalism of the previous section the approach is 
immediately generalized to any genus. We expand the matrix 
$(a_\mu)_i^j=C_1\sum (l_{\vec n})_i^j a_\mu^{\vec n}$. The constant $C_1$ is a 
normalization factor. Similarly, let us consider the gauge field for area 
preserving diffeomorphisms reduced to a single space-time point. Since the 
adjoint representation is labelled by the continuous variables $\tau,\sigma$ 
this gauge field is labelled as $A_\mu(\tau,\sigma)$. This looks like a string 
field $X_\mu(\tau,\sigma)$ defined on the Riemann surface. In order to make it 
suggestive we will label our gauge potential as $X_\mu$. For the torus it may 
be expanded as $X_\mu(\tau,\sigma)=C_2\sum a_\mu^{\vec n} exp(i\vec\sigma\cdot 
\vec n)$ where the coefficients have been labelled as $a_\mu^{\vec n}$ in 
order to establish a parallel with the coeffients of expansion for the matrix 
$(a_\mu)_i^j$ above. $C_2$ is a normalization constant which defines the 
normalization of $X_\mu$ relative to $a_\mu^{\vec n}$. Next compare the field 
strengths for the reduced SU(N) theory and the reduced area preserving gauge 
theory. The adjoint action of the area preserving diffeomorphism group (for 
any surface) is defined by the commutation rules (1), so that the adjoint 
representation is expressed by a Poisson bracket in the $\tau,\sigma$ 
variables. This yields the field strength $F_{\mu\nu}=\{X_\mu, 
X_\nu\}(\vec\sigma)\equiv \epsilon^{ij}\partial_iX_\mu\partial_jX_\nu$, which 
is nothing but the {\it string area element} for any surface. Thus, for the 
torus we have 

$$ \eqalign {
& (f_{\mu\nu})_i^j=
[a_\mu,a_\nu]_i^j = C_1^2\sum a_\mu^{\vec n}a_\nu^{\vec m}{N\over
2\pi} sin ({2\pi\over N}\vec n\times\vec m) \ (l_{\vec n+\vec m})_i^j,  \cr
&F_{\mu\nu}(\vec\sigma)=
\{X_\mu,X_\nu\}(\vec\sigma)= C_2^2\sum a_\mu^{\vec n}a_\nu^{\vec m}
(\vec n\times\vec m) \ exp(i(\vec n+\vec m)\cdot\vec\sigma).    }
\eqno(13) $$

\no
We now see that as $N\rightarrow\infty$ (for a suitable behavious of 
$a_\mu^{\vn}$ that allows the replacement $sin({2\pi\over N}\vec n\times \vec 
m)\rightarrow \vec n\times \vec m$ in the sums), the trace 
$Tr(f_{\mu\nu}f^{\mu\nu})$ will produce the same expression as the integral 
$\int d^2\sigma F_{\mu\nu}F^{\mu\nu}$  except for an overall constant. Thus, 
we find that as $N\rightarrow\infty$ the reduced action (12) takes the form 

$$        S_{red}=-{(2\pi /\Lambda)^{d-4}\over g_d^2(\Lambda)}\Big ({NC_1\over 
    2C_2\Lambda}\Big )^4 \int d^2\sigma F_{\mu\nu}F^{\mu\nu}(\vec \sigma) . 
    \eqno(14)  $$

This Lagrangian is interpreted as the square of the area element spaned by the 
string $X_\mu$. It looks different than standard string action, however if we 
write the string action in the gauge $det(\gamma)=-1$ 
with $\gamma_{ij}=\partial_iX\cdot\partial_jX$, then $S\sim \int d\tau d\sigma 
\sqrt{-\gamma}\gamma^{ij} \partial_iX\cdot\partial_jX = \int d\tau d\sigma \ 
det(\partial_iX\cdot \partial_jX)$, so (14) is seen to be identical to the 
string action semi-classically. The quantum path integral for this form of 
string theory needs to be defined by introducing a cutoff. We may regard the 
finite $N$ version of eq.(12) as the cutoff version of this string theory. 
Indeed, we have argued elsewhere [6] that this corresponds to taking a lattice 
Riemann surface with $N^2$ points. The presence of the non-trivial factor 
$f(a)$ in the measure (12) indicates that the string version of QCD differs
from standard string theory at the quantum level.

The action (12) yields the planar graphs in the standard limit (i.e. 
$N\rightarrow \infty$ first and then $\Lambda\rightarrow\infty$). 
However, we may also consider a corelated way of taking the limit in a way 
analogous to the recent double scaling limit that yields the sum over all 
genus in recent investigations of 2-dimensional gravity. Since the coupling 
constant $g_d(\Lambda)$ is really a function of $\Lambda$, this would suggest 
that in order to achieve the analog of the double scaling limit we would have 
to send $\Lambda\rightarrow\infty$ in an $N$-dependent fashion.                                        
This discussion leads to the following picture.
The path integral of the reduced model in matrix form is now expected to yield 
a path integral over the string variable $DX_\mu$ as well as a {\it sum\ over\ 
surfaces} of genus $g$, just as in string theory. 
$ \sum_g \int Dm DX_\mu f(X) \ exp ( S^{(g)}_{red}(X),$
where $f(X)$ is the measure in (12) rewritten in terms of the string variable 
$X_\mu$.

While this form is strongly reminiscent of string theory in the gauge 
$det(\gamma)=-1$, the measure in the path integral does not look quite the 
same. We really are discussing the dynamics of flux tubes of gauge theories 
rather than standard string theory. It may be 
useful to further study the consequences of (12) or (15) by using string 
techniques in order to learn non-perturbative properties of gauge theories and 
in particular of QCD in the confinement region. A place to start is d=2. This 
will be investigated in the future. 
          
Can our observations be useful in the usual string theory? In particular, can 
we use it to sum over all genus and discuss non-perturbative string physics?
To some extent this has to do with the measure being different. However, 
nothing stops us from going back to the reduced matrix action of (2) and simply 
change the measure in the path integral, so that it would be compatible with 
the required measure in string theory in the gauge $det(\gamma)=-1$. So, this 
provides us now with a matrix model which may describe the sum over all genus 
in string theory!! 

%%%%%%%%%%%%%%%%%%%%%%%%%%%%%%%%%%%%%%%%%%%%%%%%%%%%%%%%%%%%%%%%%%%%%%%
\bigskip
\no
{\bf REFERENCES}

\item {(1)} J. Hoppe, preprint PITHA 86/24; Ph.D. Thesis, M.I.T. (1982).
\item {(2)} E. Floratos and J. Illiopoulos, Phys. Lett. 201B (1988) 
237.\hfill\break 
 D. B. Fairlie, P. Fletcher and C.K. Zachos, Phys.Lett. B218.

\item {(3)} I. Bars, C.Pope and E. Sezgin, Phys. Lett. 210B (1988) 85.

\item {(4)} I. Bars, to be published.

\item {(5)} I. Bars, to be published.

\item {(6)} I. Bars, Phys.Lett.  (1990);   and ``SU(n) Gauge theory and strings on discrete Riemann surfaces, USC-90/HEP-20 (unpublished, but available through  SPIRES listing, at KEK library, 1990).

\item {(7)} For related ideas see: E.G. Floratos, J. Illiopoulos and G. 
Tiktopoulos, Phys. Lett. 217B (1989) 285. D. Fairlie and C.K. Zachos ANL-HEP-
PR-89-64.

\item {(8)} T. Eguchi and H. Kawai, Phys.Rev.Lett. 48 (1982) 47.

\item {(9)} D.Gross and Y.Kitazawa, Nucl.Phys. B206 (1982) 440.

\item {(10)} I. Bars, Phys.Lett. 116B (1982) 57.

%%%%%%%%%%%%%%%%%%%%%%%%%%%%%%%%%%%%%%%%%%%%%%%%%%%%%%%%%%%%%%%%%
\end